\documentclass[fleqn,12pt,twoside]{article}

\usepackage{espcrc1}
\usepackage{amssymb}
\usepackage[dvips]{graphicx}

\title{Statistical hadronization of charmed quarks at SPS and RHIC}
\author{A.P. Kostyuk\address{Institut f\"ur Theoretische Physik, Universit\"at  Frankfurt,
Germany\\
and\\
Bogolyubov Institute for Theoretical Physics,
Kyiv, Ukraine}}

\begin{document}
\maketitle

\begin{abstract}
Production of open and hidden charm hadrons in heavy ion collisions is 
considered within the statistical coalescence model. The charmed 
quark-antiquark pairs are assumed to be created at the initial stage 
of the reaction in hard parton collisions. The number of these pairs 
is conserved during the evolution of the system.
At the hadronization stage, the charmed (anti)quarks are 
distributed among open and hidden charm hadrons in accordance with the 
laws of statistical mechanics. 

The model is in agreement with the experimental data on J/psi to Drell-Yan
ratio in Pb+Pb collisions at SPS. This agreement can be reached only if a 
rather strong enhancement of the open charm production in central Pb+Pb 
collisions is assumed. A possible mechanism of the charm enhancement is 
discussed.

At the top RHIC energy, the model predicts  an increase of J/psi to Drell-Yan
ratio in more central nucleus-nucleus collisions with respect to less 
central ones. 
\end{abstract}

\section*{}

The interest to charmed quarks in heavy ion physics was initially 
motivated by the suggestion 
of Matsui and Satz \cite{MS}
to use charmonia as a probe of the state of matter created 
at the early stage of the collision. 

The standard picture of charmonium production in nucleus-nucleus collisions
assumes that charmonia are created exclusively at the initial stage of the 
reaction in primary nucleon-nucleon collisions. During the subsequent
evolution of the system, the number of hidden charm mesons is reduced
because of (a) absorption of pre-resonance charmonium states in the nuclei 
(normal nuclear suppression),
(b) interactions of charmonia with secondary hadrons (comovers), 
(c) dissociation of $c\bar{c}$ bound states in the deconfined medium.
It was found that the $J/\psi$ suppression with respect to
Drell-Yan muon pairs measured in proton-nucleus and 
nucleus-nucleus collisions with light projectiles 
can be explained by the normal nuclear suppression alone \cite{NA38}.
In contrast, the NA50 experiment with a heavy projectile and target
(lead-lead) revealed essentially stronger $J/\psi$ suppression for central
collisions \cite{anomalous}. 
This {\it anomalous} $J/\psi$ suppression was attributed to
formation of quark-gluon plasma \cite{evidence}.   

Despite of quite successful agreement with the $J/\psi$ data, the 
standard scenario seems to be in trouble explaining the $\psi'$ yield.
The recent lattice simulations \cite{Karsch} suggest that the temperature of 
$\psi'$ dissociation $T_d(\psi')$ lies far below the deconfinement point 
$T_c$: $T_d(\psi') \approx 0.1$--$0.2 T_c$ \cite{Satz01}. Therefore, not 
only the quark-gluon plasma, but also a hadronic co-mover medium should 
completely eliminate $\psi'$ charmonia in central Pb+Pb collisions at SPS. 
However, the experiment revealed a sizable $\psi'$ yield (see, for instance, 
\cite{Bordalo}). It was observed \cite{Shuryak} that
$\psi'$ to $J/\psi$ ratio decreases with centrality only in peripheral
lead-lead collisions, but remains approximately constant  
at sufficiently large number of participants 
$N_p \ge 100$.

A completely different picture of charmonium production was proposed
in Ref. \cite{GG}: hidden charm mesons 
are supposed to 
be created at the hadronization stage. Similar to all other hadrons,
their abundancies can be described within the thermal model \cite{thermal}.
However, production of heavy quarks in soft processes is expected
to be negligible. Most likely, 
they are produced exclusively at the hard stage. Their number, therefore, 
can, generally speaking, deviate from the thermal equilibrium value.
This fact was taken into account by the statistical coalescence model
(SCM), which was proposed in Ref.\cite{Br1}. Soon, it was realised that, 
when the average number of heavy quark pairs per collision is small 
($ \lesssim 1$), exact conservation of this pairs in the hadronization process
appears to be crucial \cite{Go:00}.   

In this talk, I present the results \cite{Ko:01,Ko:02} obtained within the 
statistical coalescence model:
\begin{itemize}
\item 
$c$ and $\bar{c}$ are created at the 
initial stage of the reaction in primary hard parton collisions;
\item
their number remains approximately unchanged during
the subsequent evolution;
\item
they are distributed over open charm hadrons and charmonia
at the hadronization stage in accordance with
laws of statistical mechanics. 
\end{itemize}

\begin{figure}[t]
\begin{minipage}[t]{7.6cm}
\includegraphics[width=7.6cm]{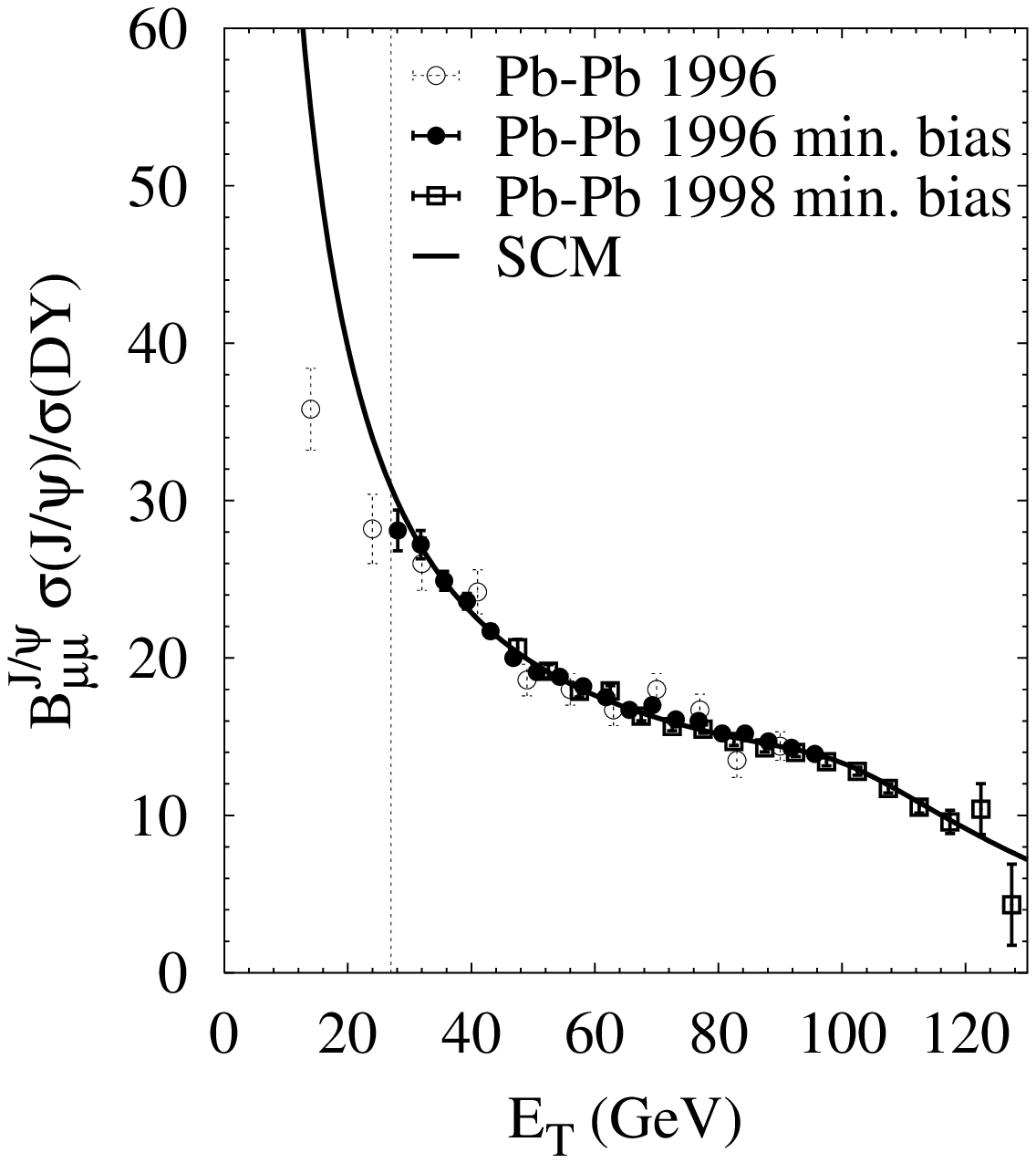} 
\mbox{}
\\[-24mm]
\mbox{}\\
\caption{The dependence of the $J/\psi$ to Drell-Yan 
ratio on the transverse energy at SPS. The vertical line shows the
boundary of the  
applicability domain of the statistical
coalescence model (SCM), see \cite{Ko:02} for details. 
\label{Jpsi_PbPb} }
\end{minipage}
\hfill
\begin{minipage}[t]{7.6cm}
\includegraphics[width=7.6cm]{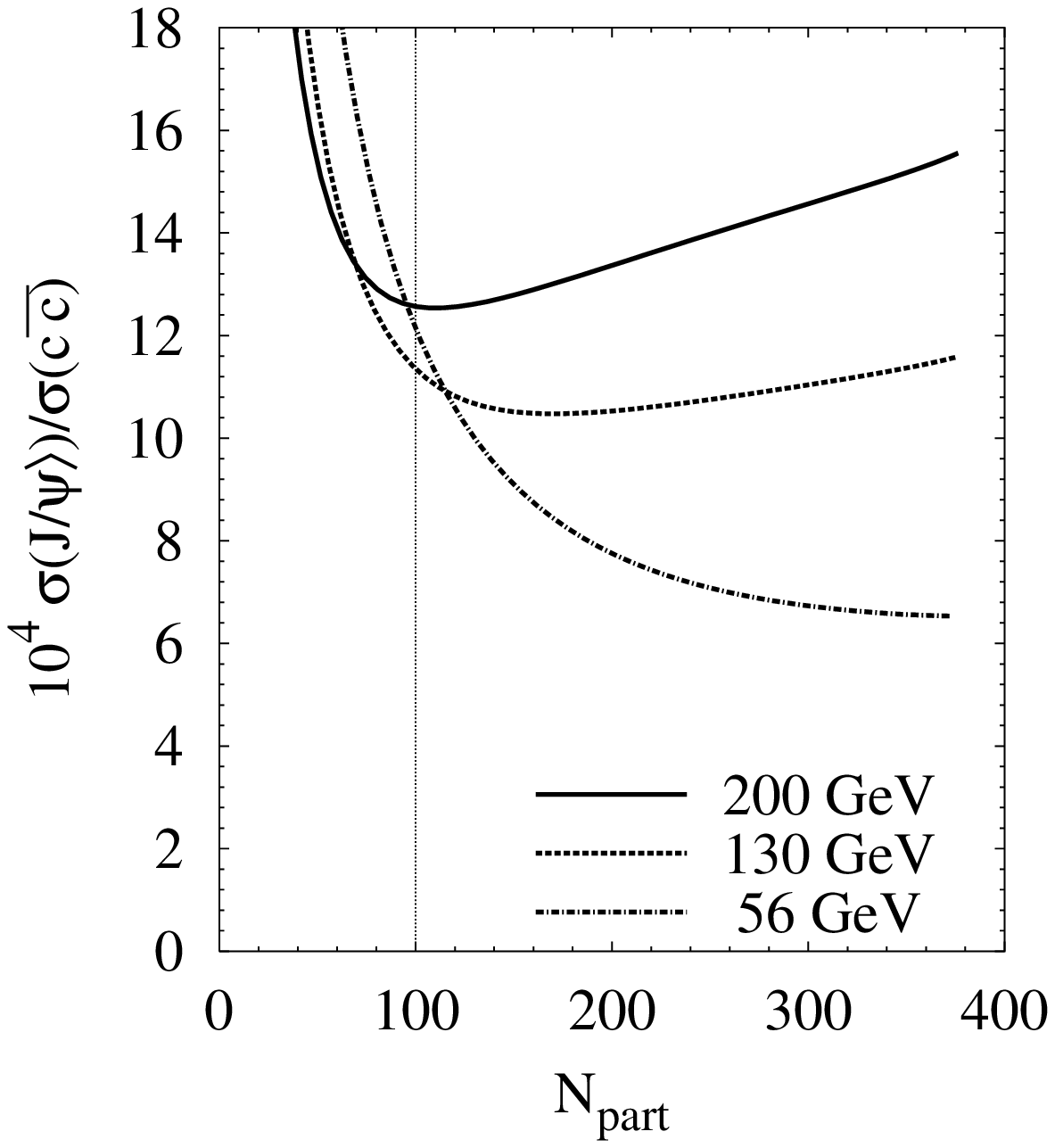} 
\mbox{}
\\[-24mm]
\mbox{}\\
\caption{The dependence of the $J/\psi$ to open charm 
ratio on the number of nucleon participants at RHIC. 
The vertical line shows the boundary of the applicability 
domain of the statistical coalescence model (SCM). 
\label{Jpsi_sl_ncc} }
\end{minipage}
\end{figure}

Within its applicability domain ($N_p \gtrsim 100$),
the model demonstrates excellent agreement  ($\chi^2/\mbox{dof} = 1.06$)
with the NA50 data on $J/\psi$ production (see Fig. \ref{Jpsi_PbPb}). 

The so-called `second threshold of anomalous $J/\psi$ suppression'
at large transverse energy ($E_T \gtrsim 100$~GeV) 
naturally appears in SCM due to two effects:
(i) fluctuations of the transverse energy at fixed number of participant 
nucleons and (ii) $E_T$-losses in the dimuon event sample with respect to
the minimum bias one  (see Ref. \cite{Ko:02} for details).

The agreement with the data is reached assuming a 
rather strong enhancement of the open charm production in central Pb+Pb 
collisions: up to a factor of about $3.5$ within the 
rapidity window of the NA50 spectrometer.
This is consistent with the indirect experimental result \cite{NA50open}.


A possible mechanism of open charm enhancement was proposed in 
Ref. \cite{hf_enh}.
Calculations in the leading order of perturbative 
quantum chromodynamics (QCD) show  that, assuming factorization (i.e.
independence of the hard QCD process on the subsequent soft stage), 
a great fraction of $c\overline{c}$ pairs is created 
with invariant masses $M_{c\overline{c}}$ below the 
corresponding meson threshold $2m_D$. In vacuum, these subthreshold pairs
would not be able to form observable hadrons. In reality, however, 
some part of these pairs may hadronize due to interaction  with spectator
partons. Nevertheless, there is no reason to assert that all subthreshold 
pairs result in final state hadrons with open or hidden charm. It is 
quite reasonable to assume that a great part of them has to 
{\it annihilate} into lighter hadrons, or, alternatively, the final state 
interactions {\it prevent} formation of such pairs (factorization is broken).

In presence of a deconfined medium (quark-gluon plasma or its precursor),
the subthreshold quark pairs can hadronize due to interactions with the 
medium, or, alternatively, the medium can modify the final state interactions
making possible creation of the subthreshold pairs and their subsequent
hadronization.
{\it  This should lead to enhanced production of the total charm 
in nucleus-nucleus collisions in comparison to the standard result obtained 
within the direct extrapolation of nucleon-nucleon data.}

We have estimated the upper bound of the open charm enhancement. 
At SPS energies, one can expect  maximal  enhancement by a factor of about 6,
which is consistent
with our above fit for the $J/\psi$ to Drell-Yan ratio.

It is interesting to note, that the statistical coalescence model does
not always lead to $J/\psi$ suppression. Predictions of the statistical 
coalescence model for RHIC are shown in Fig. \ref{Jpsi_PbPb} 
(see also Ref.\cite{RHIC}). As is seen,
$J/\psi$ suppression is still expected at the lowest RHIC energy. In contrast,
relative $J/\psi$ {\it enhancement} (e.g. growth of the $J/\psi$ to open 
charm\footnote{It should be stressed that, due to possible nuclear shadowing,
the open charm multiplicity has not to be proportional to the number of binary
nucleon-nucleon collisions. Therefore the ratio of  $J/\psi$ multiplicity to
the number of binary collisions may decrease even if $J/\psi$ to open  
charm ratio increases.}
ratio  with the number of participants at $N_p > 100$) is predicted for the
top RHIC energy. 

In conclusion, the NA50 data for
not very peripheral lead-lead collisions  are consistent 
with the following scenario:

Formation of deconfined medium leads to enhanced production of 
charmed hadrons. 
Charmonia as well as other hadrons are 
formed at the hadronization stage. 
The distribution
of charm quarks and antiquarks over open and hidden 
charm hadrons follows laws of statistical mechanics. 

A relative $J/\psi$ enhancement is expected at the top RHIC energy.

I am indebted to M.~Gorenstein, 
W.~Greiner,  L.~McLerran and H.~St\"ocker for fruitful
collaboration.  
I wish to thank A.~Andronic, F.~Becattini, P.~Bordalo, P.~Braun-Munzinger,
L.~Bravina, A.~Capella, Yu.~Dokshitzer, J.~Cleymans, W.~Florkowski, 
A.~Frawley,
L.~Grandchamp, A.~Kaidalov, B.~K\"ampfer, J.~Kapusta, F.~Karsch, 
D.~Kharzeev, V.~Koch, K.~Redlich, D.~Rischke, M.~Rosati,
H.~Satz, E.~Shuryak, 
Yu.~Sinyukov, J.~Stachel, J.~Wam\-bach, A.~Zhitnitsky and G.~Zinovjev 
for interesting discussions.


\begin{thebibliography}{99}

\bibitem{MS}
T.~Matsui and H.~Satz,
Phys.\ Lett.\ B {\bf 178} (1986) 416;\\
H.~Satz,
Rept.\ Prog.\ Phys.\  {\bf 63} (2000) 1511.

\bibitem{NA38}
M.~C.~Abreu {\it et al.},
Phys.\ Lett.\ B {\bf 466} (1999) 408.

\bibitem{anomalous}
M.~C.~Abreu {\it et al.}  [NA50 Collaboration],
Phys.\ Lett.\ B {\bf 410} (1997) 337;\\
M.~C.~Abreu {\it et al.}  [NA50 Collaboration],
Phys.\ Lett.\ B {\bf 450} (1999) 456.

\bibitem{evidence}
M.~C.~Abreu {\it et al.}  [NA50 Collaboration],
Phys.\ Lett.\ B {\bf 477} (2000) 28.

\bibitem{Karsch}
F.~Karsch, E.~Laermann and A.~Peikert,
Nucl.\ Phys.\ B {\bf 605} (2001) 579.

\bibitem{Satz01}
S.~Digal, P.~Petreczky and H.~Satz,
Phys.\ Lett.\ B {\bf 514} (2001) 57.

\bibitem{Bordalo}
M.~C.~Abreu {\it et al.}  [NA50 Collaboration],
LIP-96-04
{\it Invited talk at 26th International Symposium on Multiparticle Dynamics 
(ISMD 96), Faro, Portugal, 1-5 Sep 1996}.


\bibitem{Shuryak}
H.~Sorge, E.~Shuryak and I.~Zahed,
Phys.\ Rev.\ Lett.\  {\bf 79} (1997) 2775.

\bibitem{GG}
M.~Gazdzicki and M.~I.~Gorenstein,
Phys.\ Rev.\ Lett.\  {\bf 83} (1999) 4009.

\bibitem{thermal}
P.~Braun-Munzinger, I.~Heppe and J.~Stachel,
Phys.\ Lett.\ B {\bf 465} (1999) 15;\\
F.~Becattini, J.~Cleymans, A.~Keranen, E.~Suhonen and K.~Redlich,
Phys.\ Rev.\ C {\bf 64} (2001) 024901;\\
G.~D.~Yen and M.~I.~Gorenstein,
Phys.\ Rev.\ C {\bf 59} (1999) 2788.

\bibitem{Br1}
P.~Braun-Munzinger and J.~Stachel,
Phys.\ Lett.\ B {\bf 490} (2000) 196.

\bibitem{Go:00}
M.~I.~Gorenstein, A.~P.~Kostyuk, H.~St\"ocker and W.~Greiner,
Phys.\ Lett.\ B {\bf 509} (2001) 277;
J.\ Phys.\ G {\bf 27} (2001) L47.

\bibitem{Ko:01}
A.~P.~Kostyuk, M.~I.~Gorenstein, H.~St\"ocker and W.~Greiner,
Phys.\ Lett.\ B {\bf 531} (2002) 195.

\bibitem{Ko:02}
A.~P.~Kostyuk, M.~I.~Gorenstein, H.~St\"ocker and W.~Greiner,
J.\ Phys.\ G {\bf 28} (2002) 2297.

\bibitem{NA50open}
M.~C.~Abreu {\it et al.}  (NA38 and NA50 Collaborations) 2000
{\it Eur.\ Phys.\ J.}\ C {\bf 14} 443.

\bibitem{hf_enh}
A.~P.~Kostyuk, M.~I.~Gorenstein and W.~Greiner,
Phys.\ Lett.\ B {\bf 519} (2001) 207.

\bibitem{RHIC}
M.~I.~Gorenstein, A.~P.~Kostyuk, L.~McLerran, H.~St\"ocker and W.~Greiner,
J.\ Phys.\ G {\bf 28} (2002) 2151;\\
M.~I.~Gorenstein, A.~P.~Kostyuk, H.~St\"ocker and W.~Greiner,
Phys.\ Lett.\ B {\bf 524} (2002) 265.
\end{thebibliography}
\end{document}